# Statistical Analysis to Extract Effective Parameters on Overall Energy Consumption of Wireless Sensor Network (WSN)

Najmeh Kamyabpour, Doan B.Hoang

iNext centre for innovation in IT services and applications,
University of Technology, Sydney
najmeh|dhoang@it.uts.edu.au

*Abstract*—In this paper, we use statistical tools to analysis dependency between Wireless Sensor Network (WSN) parameters and overall Energy Consumption (EC). Our approach has two main phases: profiling, and effective parameter extraction. In former, a sensor network simulator is re-run 800 times with different values for eight WSN parameters to profile consumed energy in nodes; then in latter, three statistical analyses (p-value, linear and non-linear correlation) are applied to the outcome of profiling phase to extract the most effective parameters on WSN overall energy consumption.

## I. INTRODUCTION

Recently, wireless sensors networks (WSNs) have attracted a huge amount of attention in many fields including environment, health, disaster alert, car, building, and mining industries. They also have a great potential to revolutionize many aspects of our lives. The networking of sensors is based on the fact that sensors are most useful when they are deployed in large numbers, especially for collecting environmental map of a geographical area such as a complete building, an agriculture field, or a rain forest. Once the sensors are deployed in a sensor application, they may no longer be accessible for further physical manipulation such as fixing faulty components or changing batteries. Therefore, it becomes important to manage their Energy Consumption (EC) to increase their life time and consequently to maximize the performance of the sensor-based application. Clearly, to minimize EC of wireless sensor networks, many inter-related factors must be considered; for example, the way of sensors interconnection, sensing mechanisms, transmission mechanisms, networking protocols, topology, and routing play critical role in the overall energy consumption of a WSN system. These parts are often internally related in a complex network resulting in high complexity for network optimization.

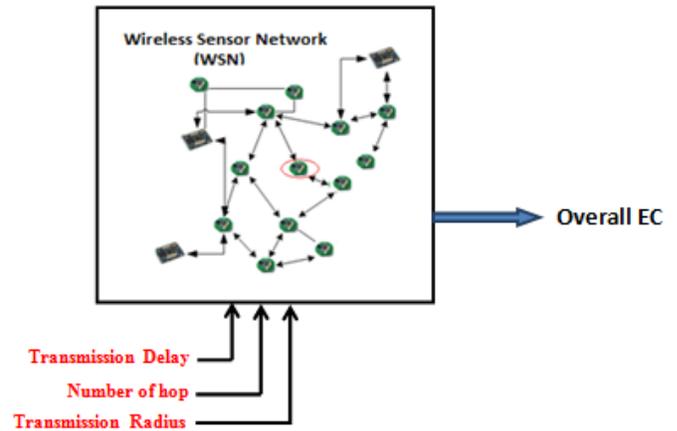

*Figure 1. Overall EC and WSN configuration parameters*

A WSN can be seen as a parametric system with several configuration parameters (figure 1). Clearly, assigning proper values for these configuration parameters can increase the performance of network and reduce overall EC. To best of our knowledge, there is no practice on study the influence of these configuration parameters on overall EC of WSN systems. The technique in this paper is our attempt to analysis and study dependency between overall EC and WSN configuration parameters, and extract effective parameters on EC regard to our simulation application. Briefly, our contributions in this paper are:

- Statistically study the influence of configuration parameters on the overall energy consumption in WSN.
- Extract most effective configuration parameters on overall EC.

The rest of the paper is structured as follows: section II highlights the energy consumption works in WSNs. Section III describes our statistical analysis to extract effective parameters in EC of WSNs, followed by experimental results and conclusion in sections IV and V, respectively.

## II. RELATED WORKS

Network architectures such as OSI and Internet are basically layer-based functional models where below layer provides services to above layer; for example, Application layer provides services to the end users. Such network architecture is often evaluated in terms of quality of service parameters such as delay, throughput, jitter, availability, reliability and even security. However, when it comes to energy consumption (EC), it becomes very difficult to evaluate and optimize EC of the network. The most difficulty is that there is no comprehensive model for a network that takes energy consumption into account. Generally, researchers focus on traditional network architectures (i.e., layer architecture) and try to minimize specific components of a single layer. They assume overall EC of a network is reduced independent from other components or layers. This is almost not an ideal situation as it is not clear how a single component fits within overall EC of an entire wireless sensor network. Almost most of current energy minimization techniques focus on send/receive data [2], while other parameters in the network are neglected. In [3-5], the proposed power consumption models use cost of sending and receiving data and furthermore deduce an upper limit for energy efficiency of a single hop distance; it considers an intermediate node between source and destination so that retransmission of data saves energy. There are some other papers to evaluate energy efficiency of wireless sensor network which mostly utilize the power consumption model in the abovementioned papers.

Due to different specification and challenges in Wireless Networks, the traditional network architecture cannot be fitted properly. Therefore, cross-layer idea has been proposed to provide flexible network architecture for Wireless Networks. The key idea in cross-layer design is to allow sharing information and consider dependency among different layers of the protocol stack [6-8]. The protocols for cross-layer design, however, are suitable and compatible for wireless networks and cannot be used for on layer-based networks. Two general examples of cross-layer design are designing of two or more joint layers, and passing of parameters between layers during run-time. however, there is no criteria to determine which layers should be combined to optimize overall EC of the network [9].

In figure 2, it was shown the overall energy consumption of WSN in term of all possible constituents and configuration parameters [1], ranging from hardware parameters (such as Dynamic-voltage frequency scaling [10, 11]) to higher level parameters. It is too probable changing value of configuration parameter decreases energy usage in one constituent but increases energy consumption in others. Therefore, it becomes important to find dependency among these configuration parameters and energy consumption of the whole network; this dependency can help researchers to focus on energy-efficient approaches in the whole system instead of optimizing energy in one specific layer.

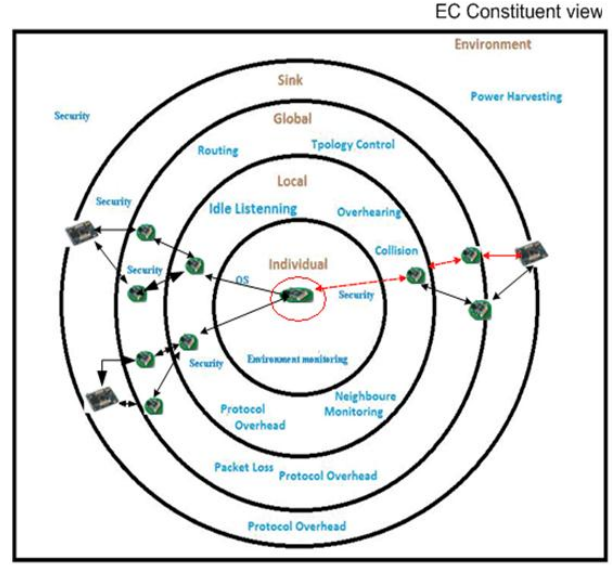

*Figure 2. possible constituents in WSN involving in overall EC[1]*

## III. EXTRACT EFFECTIVE PARAMETERS

P-value and correlation analysis are two powerful tools in statistics to analysis dependency between two random variables. To capture nonlinear relation between parameters and the output (here overall energy consumption of the network), non-linear correlation analysis is also used with p-value and linear correlation analysis. First, p-value analysis is applied to obtain the most effective parameters on overall EC. Then both linear and non-linear correlations are applied on outcome of p-value analysis to verify the result and also find the direction of dependency between these effective parameters and overall EC. In another word, p-value is specifically used to obtain the most statistically significant variables influencing the output variable while correlation is used for both influential variables and their direction effect.

a. **P-value**

TABLE 1. *P-value, linear and non-linear correlation anlayis between WSN parameters and overal energy consumption. The blue-italic-bold items show the most effective parameters based on p-value analysis*

| Parameters | P-value | Linear correlation | Non-linear correlation |
|---|---|---|---|
| **Transmission interval** | *3.7979e-005* | *-0.2842* | *-0.2474* |
| **Number of Hops** | *0.00051* | *0.2411* | *0.2247* |
| **Sensor interval** | *0.02397* | *-0.1580* | *-0.1280* |
| **Sense Radius** | *0.04933* | *0.1355* | *0.1178* |
| Network density | 0.11896 | 0.1095 | 0.0474 |
| Transmission Radius | 0.32401 | 0.0694 | 0.0693 |
| Number of Sinks | 0.42896 | 0.0557 | -0.0004 |
| Number of Neighbours | 0.44191 | 0.0541 | -0.0088 |

Making a decision about statistical significance is related to the practice of hypothesis testing. In general, the idea is to state a null hypothesis (i.e. that there is no effect) and then to see if the gathered data allows you to reject the hypothesis. In statistics, a null hypothesis ($H_0$) is tested by gathering data and then measuring how probable is the occurrence of data, assuming the null hypothesis is true. There are two cases: (1) if data is very improbable (usually defined as observed less than 5% of the time), then it can be concluded that the null hypothesis is false, and (2) if data do not contradict the null hypothesis, then no conclusion can be made. So data give insufficient evidence to make any conclusion and therefore the null hypothesis could be true or false.

In statistics, p-value is a numerical measure of the statistical significance of a hypothesis test. If the probability of effecting output regard to a specific input variable from a set of experiments is very low, we feel confident in rejecting the null hypothesis or we say the result is statistically significant. In our study, the null hypothesis ($H_0$) is that to see if k-th configuration parameter of WSN ($p_k$) has sufficient influence on the overall energy consumption($E$). If p-value between $p_k$ and $E$ is less than 0.05, then null hypothesis is rejected and $p_k$ and $E$ are highly correlated.

b. **Linear Correlation analysis**

Although p-value analysis is good enough to pick the most effective configuration parameters, it does not give any information about how the parameter and overall energy consumption are related. Precisely, p-value does not give any information about both value and sign of correlation. Therefore, correlation analysis is utilized with p-value to get more information.

By definition, correlation is a measure of dependency between two variable sequences on a scale from -1 to 1. If $\boldsymbol{p_k} = \left(p_k^{(1)}, p_k^{(2)}, \ldots, p_k^{(M)}\right)$ and $\boldsymbol{E} = \left(E^{(1)}, E^{(2)}, \ldots, E^{(M)}\right)$ are M samples of k-th parameter ($p_k$) and overall energy consumption in network, respectively, the normalized cross-correlation function between $p_k$ and $E$ can be expressed as follows:

$$Corr_{norm}(\boldsymbol{p_k}, \boldsymbol{E}) = \frac{\sum_{i=1}^{M}\left(\left(p_k^{(i)} - \mu_{p_k}\right)\left(E^{(i)} - \mu_E\right)\right)}{\sqrt{\sum_{i=1}^{N}\left(p_k^{(i)} - \mu_{p_k}\right)^2 \sum_{i=1}^{N}(E^{(i)} - \mu_E)^2}}$$

where $\mu_{p_k}$ and $\mu_E$ are the mean of $p_k$ and $E$ through M experiments, respectively. It is intuitive that a correlation value closer to 1.0 or -1.0 indicates a higher dependency between the parameter and overall energy consumption.

c. **Non-linear Correlation analysis**

As linearity of dependency between a parameter ($p_k$) and overall energy consumption ($E$) in WSN is unknown, non-linear correlation is also used, based on the study of Billings and Zhu [12, 13] and Mao and Billings[14], to capture non-linear dependency as follow:

$$Corr_{norm}^2(\boldsymbol{p_k}, \boldsymbol{E}) = \frac{\sum_{i=1}^{M}\left(\left(\left(p_k^{(i)}\right)^2 - \frac{1}{M}\sum_{j=1}^{M}\left(p_k^{(j)}\right)^2\right)\left(\left(E^{(i)}\right)^2 - \frac{1}{M}\sum_{j=1}^{M}\left(E^{(j)}\right)^2\right)\right)}{\sqrt{\sum_{i=1}^{M}\left(\left(\left(p_k^{(i)}\right)^2 - \frac{1}{M}\sum_{j=1}^{M}\left(p_k^{(j)}\right)^2\right)\right)^2 \sum_{i=1}^{M}\left(\left(\left(p_k^{(i)}\right)^2 - \frac{1}{M}\sum_{j=1}^{M}\left(E^{(j)}\right)^2\right)\right)^2}}$$

It should be noticed that $Corr_{norm}^2(p_k, E) = \pm 1$ represents perfect dependency and $Corr_{norm}^2(p_k, E) = 0$ represents strong independency between $p_k$ and $E$. Obviously for M=1, the linear correlation is calculated.

IV. EXPERIMENTAL RESULTS

A. *Experimental setting*

We have modified an open source wireless sensor network simulator to track the overall energy consumption of the whole network. The original C# code and our modified Java code of this simulator can be found at [15] and [16]. The simulator generates random events in the screen. The sensors detect events in their covered area, create packets, and send the packets to the nearest sinks in the network; these sinks are located as a group in a specific location. While a sensor has enough power it performs its defined tasks (i.e., sensing, neighbor monitoring, relay data, and routing), otherwise it stops (i.e., dies).

To test dependency between network's overall energy consumption and WSN configuration parameters and then extracting the most effective parameters, we run the simulator 800 times with different values for eight configuration parameters. These eight parameters are:
- "Sensor Interval" and "Sense Radius"
- "Transmission Radius", "Transmission interval", and "Number of Neighbours"
- "Number of Hops", "Network density", and "Number of Sinks".

Each run takes around 240 seconds and after this time the overall energy consumption of the network is captured. Then p-value, linear and non-linear correlations of the whole experiments is calculated.

## B. Results

Figure 3-a shows how overall energy consumption of the network raises by increasing number of hops in the network. ; this may be because increasing number of hops results in increasing the number of control packets and consequently consuming more energy in nodes to send data packets. This dependency can also be seen in Table 1 where number of hops parameter shows positive impact on overall EC. While increasing number of hops consumes more energy in the network, figure 3-b indicates performance slightly changes by varying the number of hops. Therefore, energy in the network can be saved by assigning the lowest possible number of hops without affecting performance.

Table 1 summarizes the result. In our simulation and based on p-value of parameters, Transmission Interval, Number of Hops, Sensor Interval, and Sense Radius are the most effective parameters on overall energy consumption of WSN due to their lowest p-values. Also, both linear and non-linear correlations show how the overall energy consumption changes by these parameters; for example more Number of Hops in network results in higher overall energy consumption (EC).

## C. Future works

Refer to our previous work in [1], there are around 34 configuration parameters which involve in overall EC of WSN. These parameters may have dependency with each other. So our next step is to find the possible dependency among these configuration parameters and reduce number of parameters. Then we would like to increase our experiment by involving more parameters in the simulation and extract the most effective parameters on overall EC. These effective parameters give a good tool to model overall EC based on these parameters; such model gives this opportunity to analysis system and estimate overall EC (with some variance) before actual running of the WSN system. The idea of modeling an output based on configuration parameters has been studied in [17, 18] and we would like to bring this experience as future work into WSN.

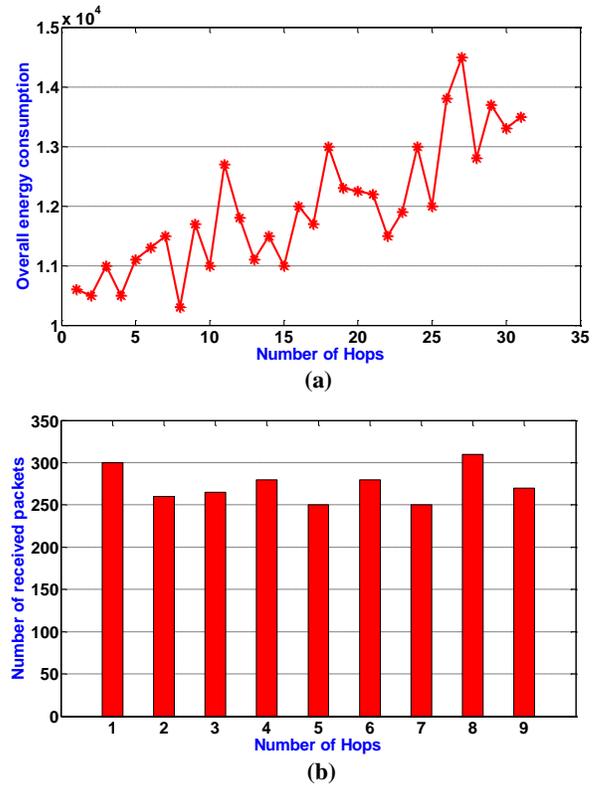

*Figure 3.* (a) Variations of overall energy consumption (EC) caused by Number of Hops (b) Network performance in term of number of received data packets in the sinks with different Number of Hops.

## V. CONCLUSION

In this paper, we utilized statistical analyses (i.e., p-value, linear correlation, and non-linear correlation) to study dependency between WSN configuration parameters and overall energy consumption of the network. Our evaluation on 800 experiments on an open source Wireless Sensor Network simulator indicated that the effectiveness of these statistical analyses resulting to find the most effective parameters on the overall energy consumption of the network.